\documentclass[twocolumn,dvipdfmx]{jpsj3-mod}
\usepackage{txfonts}
\usepackage{color}
\usepackage{graphics}
\usepackage{bm}

\title{Time-dependent renormalized quantum master equation}

\author{Akihiro Kimura and Hanayo Tsuzuku}

\inst{Department of Physics, Graduate School of Science, Nagoya University, Furo-cho, Chikusa-ku, Nagoya, 464-8602, Japan}

\abst{    Time-dependent renormalization was employed to derive a
   nonlinear 
   quantum master equation (QME), in which the dynamics of a non-equilibrium fluctuation in an irrelevant system are fed back into that of a relevant one.
   In terms of application, the nonlinear 
   QME was formulated from the viewpoint of the conventional theory for weak electronic coupling and 
analyzed numerically.
In the case of a large energy gap between sites in the relevant system and the reorganization energy, the new equation 
reproduced the results obtained through numerically exact calculations;
this behavior is in contrast with that of the conventional theory. 
}

\kword{Quantum master equation, Renormalization, Approximation method}

\begin{document}

\maketitle

\section{Introduction}
\label{}

The study of quantum dissipative systems is necessary for understanding the elementary processes of life and for the development and control of quantum devices \cite{856,May2011,Breuer-book2002}.
Many theoretical developments have been reported for such systems, and
there are two main investigative approaches: (1) nonperturbative analysis, where the system dynamics are calculated using computational resources \cite{Topaler1993285,BECK20001,Ishizaki2009,PhysRevLett.105.050404}; and (2) techniques based on analytical approximation, such as, a perturbative method with polaron representation or variational principles \cite{:/content/aip/journal/jcp/129/10/10.1063/1.2977974,:/content/aip/journal/jcp/135/3/10.1063/1.3608914,:/content/aip/journal/jcp/142/16/10.1063/1.4918736,Fujihashi2014,Kimura2014,McCutcheon2011,:/content/aip/journal/jcp/140/24/10.1063/1.4884275,doi:10.1021/jp1093492,C5CP06871A}.
  These analytical techniques limit the applicability of quantum dissipative systems, although they have the advantage that they require considerably lesser amount of computational resources.
    
To consider the dynamics of a reference system within a dissipative system, we must only construct the formalism for that system.
The projection operator technique is an easy means of deriving the closed equation of the relevant system, which is called the ``quantum master equation (QME)''. After the closed equation is derived, we can employ the appropriate approximation for the system.
For example, when the total system consists of two electronic exciton states and nuclear vibrations, there are two limiting cases. 
First, when the excitonic coupling is much smaller than the other parameters, the exciton state is localized at one site but can be transferred to another site. The transfer rate can be expressed through the F\"orster formula \cite{Forster1946} by treating the excitonic coupling as a perturbation.
Second, in the opposite limiting case, the exciton--phonon interaction is treated as the perturbative term, and Redfield theory \cite{5392713} describes the reaction dynamics between each delocalized exciton state.

The projection operator method developed by Nakajima and by Zwanzig \cite{Nakajima01121958,:/content/aip/journal/jcp/33/5/10.1063/1.1731409} reduces the information of the total system to facilitate the derivation of the QME.
However, since the conventional projection operator simply reduces the information of the irrelevant system, its dynamics are not considered by this method.
Willis and Picard proposed the time-dependent projection operator\cite{PhysRevA.9.1343,Picard1977}, which also incorporates the dynamics of the irrelevant system.
Their derived QME consists of parallel equations for the relevant and irrelevant systems and has nonlinear terms.
The merit of this approach is that it is applicable when the irrelevant system is in a non-equilibrium state.
Furthermore, Linden and May reported the derivation of a QME with nonlinear terms achieved using a time-dependent projection operator technique \cite{Linden1998}.
  In their formalism, the nonlinear QME becomes a linear one in the Markovian limit. 
Recently, Degenfeld-Schonburg and Hartmann developed the self-consistent Mori projector technique\cite{PhysRevB.89.245108}, whereby equations of reduced density operators are analytically derived for a small number of closed parallel nonlinear equations.
  This indicates that the nonlinear terms in the nonlinear equation are important for analyzing the dynamics of the reduced density matrix in a many-particle system.
However, the properties of the nonlinear term of the nonlinear QME in these theories remain unclear.

In this paper, we first utilize time-dependent projection operator techniques to derive a conventional nonlinear QME, in the manner of Shibata and Hashitsume\cite{Shibata1979}. 
Next, we apply time-dependent renormalization to this derived equation, drastically approximate the new equation to an appropriate perturbation order, and apply it to the weak electronic coupling model.
Furthermore, we present the numerical results obtained with this new equation and compare them with those produced by the conventional theory and by a numerically exact calculation.
Finally, we discuss the various properties of the new theory.

\section{Derivation of master equations}

We start by defining the Hamiltonian and its Liouville operator.
The total system is constructed as the sum of ``relevant'' (matter) and ``irrelevant'' (field) parts that can interact with each other.
Hence, the total Hamiltonian is expressed as
\begin{align}
  H &= H_m + H_f +  H_{m-f}, 
      \label{eq:Hamiltonian}
\end{align}
where, for example,  
$H_m$ is the electronic Hamiltonian for the relevant (matter) system, $H_f$ is the irrelevant (field) Hamiltonian describing the environment nuclear vibration (phonon field), and $ H_{m-f}$ is the interaction Hamiltonian between the relevant and irrelevant systems. 
Based on the Hamiltonians, we define each eigenvector as $|\phi_i^f \rangle$ ($|\phi_j^m \rangle $) of the $i$th ($j$th) quantum number for the relevant (irrelevant) Hamiltonian.

For perturbative approximation, we assume that the total Hamiltonian of Eq. (\ref{eq:Hamiltonian}) can be divided into two terms as follows:
\begin{align}
  H &= H^0 +  H'
      ,
\end{align}
where $H^0$ is the nonperturbative Hamiltonian, and $H'$ is the perturbative Hamiltonian.
The definition of the Liouville operator is completely analogous.
Based on the Liouville operator, the Liouville--von Neumann equation in the interaction representation can be written as
\begin{align}
  i\hbar \frac{\partial W_I(t)}{\partial t} =  L'_I(t) W_I(t) 
  \equiv [H'_I(t), W_I(t)],
  \label{eq:Liouville_eq}
\end{align}
where $W_I(t)$ is the density operator in the interaction representation and $L'_I(t)$ is the interaction representation of $L'$, defined as $U_0^{\dag}(t) L' \equiv e^{iL^0t/\hbar} L'$.

\begin{figure}
  \centering
    \includegraphics*[width=8cm]{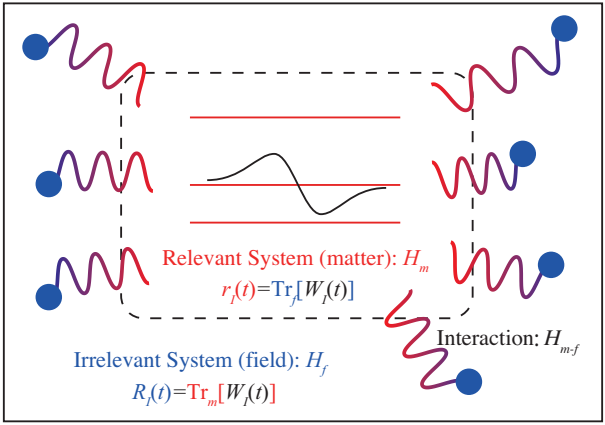}
    \caption{Schematic view of relevant (red) and irrelevant (blue) systems. The red lines and blue circles represent the possible states of the relevant and irrelevant systems, respectively. These two systems interact through $H_{m-f}$, and this interaction is represented by the wavy lines.}
    \label{fig:projection-operator}
  \end{figure}

To derive the QME,
the time-dependent projection operator\cite{PhysRevA.9.1343} is defined as
\begin{align}
  P(t) = R_I(t) \mbox{Tr}_f + r_I(t) \mbox{Tr}_m - R_I(t) r_I(t) \mbox{Tr},
  \label{eq:P(t)}
\end{align}
where $\mbox{Tr}_f$ and $\mbox{Tr}_m$ are the exclusive trace operators for the irrelevant and relevant systems, respectively  (see Fig. \ref{fig:projection-operator}),
and $\mbox{Tr}$ is defined as $\mbox{Tr}_f \mbox{Tr}_m$. Finally,
$r_I(t)$ and $R_I(t)$ are the reduced density operators for the relevant and irrelevant systems, respectively, defined as
\begin{align}
  r_I(t) = \mbox{Tr}_f [ W_I (t) ] = \sum_i \langle \phi^f_i| W_I(t) | \phi^f_i \rangle ,
  \label{eq:def_of_r} \\
  R_I(t) = \mbox{Tr}_m [W_I (t) ] = \sum_j \langle \phi^m_j| W_I(t) | \phi^m_j \rangle .
  \label{eq:def_of_R}
\end{align}
The time-dependent projection operator satisfies the following relations:
$[P(t), \partial / \partial t] \equiv -\dot{P}(t)$, $\dot{P}(t) W_I(t) = 0$,
$P(t_1)P(t_2) \equiv P(t_1)$, $Q(t_1) Q(t_2) \equiv Q(t_2)$, and $P(t_1) Q(t_2) \equiv 0$,
where we define 
$Q(t) = 1- P(t)  $.

\subsection{Time-convolutionless QME}

According to the derivation by Shibata and Hashitsume\cite{Shibata1979}, applying the time-dependent projection operator of Eq. (\ref{eq:P(t)}) to the Liouville--von Neumann equation of Eq. (\ref{eq:Liouville_eq}) under the initial condition of $Q(0)W_I(0) = 0$, we obtain the following equation for the reduced density matrix in the interaction representation:
\begin{align}
  i\hbar \frac{\partial P(t)W_I(t)}{dt}
  &=
  P(t)L'_I(t)
  \Theta(t)
    P(t) W_I(t)
    ,
  \label{eq:eq_of_P(t)W(t)}
\end{align}
where
\begin{align}
   &        U^Q(t)
  =
  \exp _+ \left[ -\frac{i}{\hbar}\int_0^t ds Q(s) L'_I(s) \right],
  \\
  & \Theta (t) =
    \left[
  1
  +
  \frac{i}{\hbar}
  \int_0^t ds
  U^Q(s)  
  Q(s) L'_I(s)
  P(s) G(t,s)
    \right]^{-1}
    ,
           \\
   &G(t,s) = \exp_- \left[ -\frac{i}{\hbar} \int_s^t ds' L'_I(s') \right]
     .
\end{align}
Taking the $f$ or $m$ traces of Eq. (\ref{eq:eq_of_P(t)W(t)}), and using the definition of Eqs. (\ref{eq:def_of_r}) and (\ref{eq:def_of_R}), we derive the time-convolutionless QME in the interaction representation as follows:
\begin{align}
&  i\hbar
  \frac{\partial r_I(t)}{\partial t} 
  =
  \mbox{Tr}_f
  [
  L'_I(t)
  \Theta(t)
  r_I(t) R_I(t) 
  ],
  \label{eq:exact_r}
  \\
&  i\hbar
  \frac{\partial R_I(t)}{\partial t} 
  =
  \mbox{Tr}_m
  [
  L'_I(t)
  \Theta(t)
  r_I(t) R_I(t) 
  ].
  \label{eq:exact_R}
\end{align}

\section{Time-dependent renormalized nonlinear QME}

\subsection{General expression}
\label{section:3.1}

To renormalize 
the irrelevant dynamics of Eq. (\ref{eq:exact_R}) into the relevant dynamics of Eq. (\ref{eq:exact_r}), 
we respectively define the renormalized nonperturbative and perturbative Hamiltonians as
\begin{align}
  &  H^{0r}(t)
    =
    H^0
    +
    C
    h_m(t),
    \label{eq:H0r}
\\
&  H'^{r}(t)
  =
  H'
    -
     C
     h_m(t),
     \label{eq:H'r}
\end{align}
where $C$ is an arbitrary real variable, and $h_m(t)$ is defined as the time-dependent operator to renormalize the dynamics of the irrelevant system into that of the relevant system.
  Eq. (\ref{eq:exact_r}) is an equation of the reduced density matrix $  r_{I}(t) $ with respect to the relevant system. However, the reduced density matrix $ R_{I}(t) $
 of the irrelevant system is also included in this equation. Therefore, it is difficult to obtain the analytical/numerical results of $  r_{I}(t) $ exclusively from Eq. (\ref{eq:exact_r}), unless we solve the time dependency of   $ R_{I}(t) $. For this purpose, a renormalization operation is performed such that the renormalized $ R_{I}(t) $
 always has an analytical solution that does not depend on time, yet is necessary to define the initial condition.
 Based on this strategy, we introduced the factor $C$ to consider the time-dependency of the renormalized $R_{I}(t)$ perturbatively fluctuating in the renormalized state.
   Apparently, Eqs. (\ref{eq:H0r}) and (\ref{eq:H'r}) are not renormalized when $C=0$ holds. 
  Thus, we define the renormalized Hamiltonian such that the reduced density matrix of the irrelevant system is independent of time when $C=1$ holds.
The definitions for the renormalized nonperturbative and perturbative Liouville operators are completely analogous.

  In the same manner as described in the previous section,
we can derive the time-dependent renormalized QME in the interaction representation as
\begin{align}
&  i\hbar
  \frac{\partial r_{\mathcal{I}}(t)}{\partial t} 
  =
  \mbox{Tr}_f
  [
  L_{\mathcal{I}}'^r(t)
  \Theta^r(t)
  r_{\mathcal{I}}(t) R_{\mathcal{I}}(t) 
  ],
  \label{eq:exact_renorm_r}
  \\
&  i\hbar
  \frac{\partial R_{\mathcal{I}}(t)}{\partial t} 
  =
  \mbox{Tr}_m
  [
  L_{\mathcal{I}}'^r(t)
  \Theta^r(t)
  r_{\mathcal{I}}(t) R_{\mathcal{I}}(t) 
       ]
       ,
  \label{eq:exact_renorm_R}
\end{align}
where
\begin{align}
   &        U^{rQ}(t)
  =
  \exp _+ \left[ -\frac{i}{\hbar}\int_0^t ds Q(s) L_{\mathcal{I}}'^r(s) \right],
  \\
  & \Theta^r (t) =
    \left[
  1
  +
  \frac{i}{\hbar}
  \int_0^t ds
  U^{rQ}(s)  
  Q(s) L_{\mathcal{I}}'^r(s)
  P(s) G^r(t,s)
    \right]^{-1}
    ,
    \label{eq:renorm_theta}
  \\
   &G^r(t,s) = \exp_- \left[ -\frac{i}{\hbar} \int_s^t ds' L_{\mathcal{I}}'^r(s') \right]
     .
\end{align}
Here, we assume the initial condition of $Q(0)r_{\mathcal{I}}(0) = 0$, 
and
introduce the renormalized perturbative Liouvillian $L_{\mathcal{I}}'^r(t)$ in the interaction representation as
\begin{align}
  L_{\mathcal{I}}'^{r}(t) A &= [u^{\dag}(t) H'^r(t) u(t),A]
,
\end{align}
where we introduce the renormalized nonperturbative propagator $u(t)$ as
\begin{align}
  u(t)
  =
  \exp _+
  \left[
  -\frac{i}{\hbar}
  \int_0^t ds
  H^{0r}(s)
  \right]
  .
  \label{eq:non-p_propagator}
\end{align}
In the interaction representation, the index $\mathcal{I}$ differs from the index $I$ in the previous section because of the renormalization.
It should be noticed that, when $C=0$ holds, Eqs. (\ref{eq:exact_renorm_r}) and (\ref{eq:exact_renorm_R}) are equivalent to Eqs. (\ref{eq:exact_r}) and (\ref{eq:exact_R}), respectively.

Because of the relation $\mbox{Tr}_m[A R_{\mathcal{I}}(t)] = \mbox{Tr}_m[A ]R_{\mathcal{I}}(t)$ for any superoperator $A$, Eq. (\ref{eq:exact_renorm_R}) is expressed by the commutator as
\begin{align}
  &i\hbar  \frac{\partial R_{\mathcal{I}}(t)}{dt}
    =
    [
    \mbox{Tr}_m
    [
    u^{\dag}(t)
    (
      H'
    -
    C h_m(t)
    )
    u(t)
    \Theta^r(t)
    r_{\mathcal{I}}(t)
    ]
    ,
    R_{\mathcal{I}}(t)
    ]
    .
\end{align}
We define the operator $h_m(t)$ and its interaction representation $h'_m(t)$
to satisfy $\partial_t R_{\mathcal{I}}(t) = 0$ when $C=1$ holds, as
\begin{align}
  h_m(t) &= u(t) h'_m(t) u^{\dag}(t)
           ,
           \\
  h'_m(t) &= 
      \mbox{Tr}_m
    [
  H_{\mathcal{I}}'^r(t)
 \Theta^r(t)
    r_{\mathcal{I}}(t)
  ]
  \frac{1}
  {
    \mbox{Tr}_m
    [
  \Theta^r(t)
  r_{\mathcal{I}}(t)
  ]
            }
            ,
            \label{eq:selfconsistent_hm}
\end{align}
where $1/    \mbox{Tr}_m
    [
 \Theta^r(t)
  r_{\mathcal{I}}(t)
  ]
$
is the inverse superoperator of $\mbox{Tr}_m
    [
 \Theta^r(t)
  r_{\mathcal{I}}(t)
  ]$.
    Eq. (\ref{eq:selfconsistent_hm}) is expressed by the weight operator of $\Theta^r(t) r_{\mathcal{I}}(t)$ as it averages $  H_{\mathcal{I}}'^r(t)$. Because the operator $\Theta^r(t) r_{\mathcal{I}}(t)$ includes the operator   $h'_m(t)$, Eq. (\ref{eq:selfconsistent_hm}) is a self-consistent equation without any approximations. The physical significance of $h'_m(t)$, is, therefore, the self-energy for the relevant system.

  In the following, we assume that
the initial condition of
the renormalized density operator for the irrelevant system $R_{\mathcal{I}}(0)$ should correspond to the thermal equilibrium state as 
$R_{\mathcal{I}}(0)
=
\rho_B=e^{-\beta H_f}/\mbox{Tr}_f[e^{-\beta H_f}]$.
The superoperator in the Liouville space for $h'_m(t)$ is defined as
$  l'_m(t) A = [
h'_m(t), A],
$ 
for any operator $A$.
Using the relation $\mbox{Tr}_f[l'_m(t) A] = 0$, 
$\mbox{Tr}_f[L_{\mathcal{I}}'^r(t) A]$ can be expressed as $\mbox{Tr}_f[L_{\mathcal{I}}'(t) A] $. Consequently, the equation $r_{\mathcal{I}}(t)$ is ultimately re-expressed as follows:
\begin{align}
&  i\hbar
  \frac{\partial r_{\mathcal{I}}(t)}{\partial t} 
  =
  \mbox{Tr}_f
  [
  L_{\mathcal{I}}'(t)
  \Theta^r(t)
  r_{\mathcal{I}}(t) R_{\mathcal{I}}(t) 
                ]
                .
  \label{eq:exact_renorm_r-2}
\end{align}
\subsection{Perturbative approximation}

In this section, we drastically approximate the renormalized equation of $r_{\mathcal{I}}(t)$ in Eq. (\ref{eq:exact_renorm_r-2}) to the nonlinear QME with the inclusion of the correction terms for the conventional QME. For this purpose, we first perturbatively approximate the operator $\Theta^r(t)$ of Eq. (\ref{eq:renorm_theta}) as
\begin{align}
  \Theta^r(t) A
&
  \simeq
A  -  \frac{i}{\hbar}
    \int_0^t
    ds
      Q(s)[(H'_{\mathcal{I}}(s) -
                  Ch'_m(s)),P(s) A]
                  ,
      \label{eq:approx-theta}
\end{align}
where we approximate the self-consistent equation of $h'_m(t)$ in Eq. (\ref{eq:selfconsistent_hm}) by the lowest order of $H'_{\mathcal{I}}(t)$ as
\begin{align}
  h'_m(t)
  =
    \mbox{Tr}_m[
  u^{\dag}(t) H' u(t)
  r_{\mathcal{I}}(t) 
  ]
  .
  \label{eq:h_m}
\end{align}
Hence, by inserting Eq. (\ref{eq:approx-theta}) into Eq. (\ref{eq:exact_renorm_r-2}), the second-order perturbative QME for $r_{\mathcal{I}}(t)$ 
is re-expressed as
\begin{align}
  i\hbar
  \frac{\partial r_{\mathcal{I}}(t)}{\partial t} 
  &
    =
  \mbox{Tr}_f
  [
                L_{\mathcal{I}}'(t)
                r_{\mathcal{I}}(t)
                R_{\mathcal{I}}(t) 
                ]
                \notag \\
&  -
  \frac{i}{\hbar}
  \int_0^t ds
    \mbox{Tr}_f
  [
  L_{\mathcal{I}}'(t)
  Q(s)
  L_{\mathcal{I}}'^{r}(s)
  r_{\mathcal{I}}( t ) R_{\mathcal{I}}(0) 
                            ],
                            \label{eq:eq_of_r-29}
\end{align}
where we replace the operator $R_{\mathcal{I}}(t)$ in the second term on the r.h.s. in Eq. (\ref{eq:eq_of_r-29}) by $R(0)$, because the perturbative order in its term is second.

  Let us consider that the renormalized nonperturbative propagator $u(t)$ in $L_{\mathcal{I}}'^{r}(t)$ in Eq. (\ref{eq:eq_of_r-29}) is perturbatively expanded and approximated by first order. 
\begin{align}
  L_{\mathcal{I}}'^r(t)              & \simeq
                                       L_I'^r(t)
             +\frac{i}{\hbar} \int_0^t ds
                          C
                          l_m'(s)
             L_I'^r(t).
                          \label{eq:Approx_LI}
\end{align}
  In the same way, we perturbatively expand the operator $R_I(t)$ by first-order perturbation.
Hence, we obtain 
\begin{align}
  &  
    R_{\mathcal{I}}(t) 
    \simeq
    \rho_B
    -
    \frac{i}{\hbar}
    \int_0^t
    ds
    \mbox{Tr}_m
  [
  (L_I'(s) - Cl'_m(s))
  r_{\mathcal{I}}(s) \rho_B 
  ].
    \label{eq:Approx_RI}
\end{align}

By inserting Eqs. (\ref{eq:Approx_LI}) and (\ref{eq:Approx_RI}) into Eq. (\ref{eq:eq_of_r-29}),
  we obtain the equation for $r_{\mathcal{I}}(t)$ as
\begin{align}
  &
  i\hbar
  \frac{\partial r_{\mathcal{I}}(t)}{\partial t} 
    =
  \mbox{Tr}_f
  [
                L_I'(t)
                r_{\mathcal{I}}(t)
                \rho_B 
                ]
                \notag \\
&  -
  \frac{i}{\hbar}
  \int_0^t ds
    \mbox{Tr}_f
  [
                            L_I'(t)
                            (
                            1
                            -
                            (
                            \rho_B
                            \mbox{Tr}_f
                            +
                            r_{\mathcal{I}}(s)
                            \mbox{Tr}_m
                            )
                            )
                            L_I'(s)
                            r_{\mathcal{I}}(t) \rho_B
                            ]
                \notag \\
  &
    -
    (1-C)
    \frac{i}{\hbar}
    \int_0^t
    ds
    \mbox{Tr}_f
    [
    L_I'(t)
    r_{\mathcal{I}}(t) 
    \mbox{Tr}_m
    [
    L_I'(s)
    r_{\mathcal{I}}(s) \rho_B 
    ]]
    .
    \label{eq:final-general-express}
\end{align}
In the Markov approximation, we replace the time dependency in the nonlinear QME in Eq. (\ref{eq:final-general-express})
$r_{\mathcal{I}}(s)$ with $r_{\mathcal{I}}(t)$
, and   then replace the variable $s$ with $t-s$ in the time integration.
Ultimately, 
we obtain the second-order nonlinear QME as
\begin{align}
  i\hbar
  &
  \frac{\partial r_{\mathcal{I}}(t)}{\partial t} 
    =
  \mbox{Tr}_f
  [
                L_I'(t)
                r_{\mathcal{I}}(t)
                \rho_B 
                ]
                \notag \\
 &
    -
  \frac{i}{\hbar}
   \int_0^{t} ds
    \mbox{Tr}_f
[
                            L_I'(t)
                            (
                            1
                            -
                            \rho_B
                            \mbox{Tr}_f
                            )
                            L_I'( t-s )
                            r_{\mathcal{I}}(t) \rho_B
                            ]
                \notag \\
&  {\color{red} + }
                            C
  \frac{i}{\hbar}
                            \int_0^{t} ds
    \mbox{Tr}_f
[
                            L_I'(t)
                            r_{\mathcal{I}}(t)
                            \mbox{Tr}_m
                            L_I'( t-s )
                            r_{\mathcal{I}}(t) \rho_B
                            ].
    \label{eq:final-markovian-express}
\end{align}
It should be noted that Eq. (\ref{eq:final-markovian-express}) becomes a conventional linear QME when the parameter C is 0.
On the other hand, because $R_{\mathcal{I}}(t)$ 
is independent of time when the parameter C is 1, Eq. (\ref{eq:final-markovian-express}) becomes a nonlinear QME.

When we perform the time integration into infinity in Markov limit, we obtain the following equation:
  \begin{align}
  i\hbar
  &
  \frac{\partial r_{\mathcal{I}}(t)}{\partial t} 
    =
  \mbox{Tr}_f
  [
                L_I'(t)
                r_{\mathcal{I}}(t)
                \rho_B 
                ]
                \notag \\
 &
    -
  \frac{i}{\hbar}
   \int_0^{\infty} ds
    \mbox{Tr}_f
[
                            L_I'(t)
                            (
                            1
                            -
                            \rho_B
                            \mbox{Tr}_f
                            )
                            L_I'(t-s)
                            r_{\mathcal{I}}(t) \rho_B
                            ]
                \notag \\
&  {\color{red} + }
                            C
  \frac{i}{\hbar}
  \int_0^{\infty} ds
    \mbox{Tr}_f
[
                            L_I'(t)
                            r_{\mathcal{I}}(t)
                            \mbox{Tr}_m
                            L_I'(t-s)
                            r_{\mathcal{I}}(t) \rho_B
                            ],
    \label{eq:final-Born-markovian-express}
  \end{align}
  which is a type of Born-Markov approximation to derive the conventional QME.

\section{Analytical application} 

As an analytical application, we used the following spin-boson Hamiltonian for a two-electronic state system:
\begin{align}
      H^0 &= \sum_{i = d, a} H_i 
            |i\rangle \langle i|
   ,
            \label{eq:Hamiltonian-H0}
  \\
      H_i &=  E_i 
            + H_i^{e-p}
   ,
    \label{eq:Hamiltonian_Hi}
  \\
  H' &=
      \sum_{i \neq j}^{d,a} V_{ij} [|i\rangle \langle j| + |j\rangle \langle i|]
       \label{eq:H'-forster}
       ,
  \\
  H_i^{e-p} &=\sum_{k}  \hbar \omega_{k} [ b_k^{\dag} b_k
              +  g_{ik} (b_k^{\dag} + b_k) ],
              \label{eq:Hamiltonian_Hep}
\end{align}
where $E_i$ is the energy of the electronic excited state $|i\rangle$ in the $i$th site, and $V_{ij}$ denotes the exciton coupling strength between the $i$th and $j$th sites.
In addition to the electronic states, we considered the phonon bath for nuclear vibration only: $b_k^{\dag}$ ($b_k$) is the creation (annihilation) operator for the $k$th phonon mode, and its frequency is $\omega_k$.
Finally, $g_{ik}$ is the exciton--phonon coupling strength for the $i$th excited state and the $k$th phonon mode.

In the following section, we numerically analyze the property of
  Eq. (\ref{eq:final-Born-markovian-express}) under secular approximation.
  The expression of the elements of the reduced density matrix $r_{\mathcal{I}}(t)$ is given in the Appendix.
  Hence, according to Eq. (\ref{eq:NLQME_for_summarized}), 
  we ultimately obtain the nonlinear QME for the weak electronic coupling limit as follows:
\begin{align}
\frac{\partial r_{dd}(t) }{\partial t} 
  &= k_{da}r_{aa}(t) - k_{ad} r_{dd}(t)
     - C (k_{da} - k_{ad}) r_{ad}(t) r_{da}(t)
    ,
\label{eq:nlqme-for-forster1}
  \\
\frac{\partial r_{da}(t) }{\partial t} 
  &= -k_{ad} r_{da}(t) + C K_{da}(r_{aa}(t) - r_{dd}(t) ) r_{da}(t)
    ,
                                       \label{eq:nlqme-for-forster2}
\end{align}
where
$k_{da}$ and $K_{da}$ are respectively expressed as
\begin{align}
  k_{ad}
&= \frac{V_{ad}^2  }{\hbar^2} \lim_{t \to \infty} \int_0^{t} ds 
[
\langle
e^{iH_dt/\hbar}e^{-iH_a s/\hbar}e^{-iH_d (t-s)/\hbar}
\rangle_B
\notag \\
  & 
+
\langle
e^{iH_d(t-s)/\hbar}e^{iH_a s/\hbar}e^{-iH_dt/\hbar}
\rangle_B
    ]
    ,
        \label{eq:kda-rate-constant}
  \\
  K_{da} 
&= \frac{V_{ad}^2  }{\hbar^2} \lim_{t \to \infty} \int_0^{t} ds 
[
\langle
e^{iH_dt/\hbar}e^{-iH_a s/\hbar}e^{-iH_d (t-s)/\hbar}
\rangle_B
\notag \\
  & 
-
\langle
e^{iH_a (t-s)/\hbar}e^{iH_d s/\hbar}e^{-iH_at/\hbar}
\rangle_B
    ]
    ,
\label{eq:Kad-org}
\end{align}
where the bracket $\langle \cdots \rangle _B$ is defined as $\mbox{Tr}_f [\cdots \rho_B]$, $\rho_B$ is $e^{-\beta H_B}/\mbox{Tr}_f[e^{-\beta H_B}]$, and $H_B$ is bath Hamiltonian defined as $\sum_k \hbar \omega_k b_k^{\dag} b_k$.

For Eqs. (\ref{eq:kda-rate-constant}) and (\ref{eq:Kad-org}), we assume that the vibrational relaxation is very fast, and that the distribution expressed by $\rho_B$ may represent a thermal equilibrium state in which the excitation almost localizes at each electronic state as $e^{-iH_i t/\hbar} \rho_B e^{iH_i t/ \hbar} \simeq \rho_i$, 
where $\rho_i = e^{-\beta H_i} / \mbox{Tr}_f [e^{-\beta H_i} ]$  for $i=d$ or $a$. With this assumption, we approximate Eqs. (\ref{eq:kda-rate-constant}) and (\ref{eq:Kad-org}) as follows:
\begin{align}
  k_{ad}
  &=\frac{V_{ad}^2}{\hbar^2} \int_{-\infty}^{\infty} d\tau \langle e^{iH_a\tau/\hbar} e^{-iH_d\tau/\hbar} \rangle_d
    \label{eq:kda-rate-constant2}
    ,
     \\
     K_{da} &= \frac{V_{ad}^2}{\hbar^2} \int_0^{\infty} d\tau
           [
           \langle e^{iH_d\tau/\hbar}e^{-iH_a\tau/\hbar}\rangle_d
           -
           \langle e^{-iH_a\tau/\hbar}e^{iH_d\tau/\hbar}\rangle_a
           ]
           ,
    \label{eq:correlation_function_for_forster}
\end{align}
where we define the brackets as $\langle \dots \rangle _i = \mbox{Tr}_f[\dots \rho_i]$.
According to the definition of the Hamiltonian in Eq. (\ref{eq:Hamiltonian_Hi}),
the correlation function in the integrand of Eqs. (\ref{eq:kda-rate-constant2}) and (\ref{eq:correlation_function_for_forster}) is expressed as
\begin{align}
  &
    \langle
    e^{iH_a t /\hbar}
    e^{-iH_d t /\hbar}
    \rangle_d
    =
    e^{
    i(E_a - E_d + \lambda_d + \lambda_a) t /\hbar
    -g_d(t)-g_a(t)
    }
    ,
\end{align}
where
\begin{align}
  g_i(t)
  &
    =
    \int_0^{t} dt_1
    \int_0^{t_1} dt_2
    \int_{-\infty}^{\infty}
    d \omega
    J_i(\omega)
    (n(\omega)+1)
    e^{ -i\omega t_2}
    ,
\end{align}
for $i=d$ or $a$.
Here, $J_{i}(\omega)$ is the density of states with the exciton--phonon interaction
\begin{align}
  J_i(\omega) = 
  \sum_k
  \omega^2
  g_{ik}^2
  \delta (\omega - \omega_k)
  ,
\end{align}
and it has the property $J_i(-\omega) = -J_i(\omega)$\cite{Yang2002,doi:10.1063/1.3155214}. Here, $n(\omega)$ is the Bose--Einstein distribution function, and 
$  \lambda_i$ is the reorganization energy of the $i$th state defined as $
\hbar  \int_0^{\infty} d\omega
  J_i(\omega)/\omega
$.

The factors  $k_{ad}$ and $k_{da}$ are equivalent to the rate constant expression in F\"orster theory\cite{Forster1946}. The factor $K_{da}$ is a constant that is involved in nonlinear effects and the integrand has the form of
  the difference between the time correlation functions of forward and backward reactions.
Owing to the expressions in Eqs. (\ref{eq:nlqme-for-forster1}) and (\ref{eq:nlqme-for-forster2}), if the off-diagonal elements $r_{ad}(t)$ and $r_{da}(t)$ of the density operator corresponding to coherence are zero as in the initial condition, the effect of the nonlinear term does not appear.
In addition, even when the values of the diagonal elements of the density operators $r_{dd}(t) $ and $r_{aa}(t) $ are nearly equivalent, the effect of the nonlinear term in Eq. (\ref{eq:nlqme-for-forster2}) is small. Taking this aspect into consideration, the next section describes numerical calculations of the nonlinear QME, insofar as the initial condition has some electronic coherence in the relevant system.

\section{Numerical analysis}

For the numerical analysis, we assumed that $J_i(\omega)$
is independent of the state $|i\rangle$ by denoting it as $ J(\omega)$,
and we used the overdamped Brownian oscillator model given by
\begin{align}
  J(\omega) 
  = \frac{2}{\pi} \frac{\lambda \gamma \omega}{\hbar^2 \omega^2 + \gamma^2}
.
\end{align}

Furthermore, we used the following parameters: exciton coupling $V_{ad} = 20$ cm$^{-1}$, reorganization energy $\lambda = 200$ cm$^{-1}$
(which is 10 times larger than the value of $V_{ad}$) to satisfy the perturbative approximation), cutoff energy $\gamma = 500$ cm$^{-1}$ (which is proportional to the inverse of the relaxation time $\tau_B$ in the nuclear fluctuations at about 10 fs) to satisfy the Markov approximation, and temperature $T = 300$ K. The energy gap $E_d - E_a$ between the donor and acceptor was 500 cm$^{-1}$.
  It was apparent that the value of $V_{ad}$ was smaller than the other parameters.
  In the site representation, the initial conditions for the numerical analysis were $r_{dd}(0) = 0.9$, $r_{aa}(0) =0.1$, and $r_{da}(0) = r_{ad}(0) = \sqrt{r_{dd}(0) r_{aa}(0)}$.

\begin{figure}
   \centering
     \includegraphics[width=7cm]{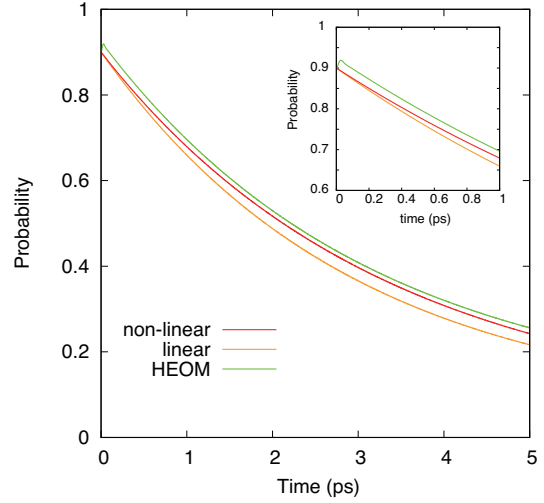}    
     \caption{Time profile of probability $r_{dd}(t)$ calculated using the proposed theory (red), linear QME (orange), and HEOM (green) in the case where $V_{da} = 20$ cm$^{-1}$, $\lambda = 200$ cm$^{-1}$, $\gamma = 500$ cm$^{-1}$, $T = 300$ K, and $E_d - E_a = 500$ cm$^{-1}$.
         The inset shows the time profile of the same $r_{dd}(t)$ over a short period of time. 
     }
     \label{fig:result-fig}
   \end{figure}

  We numerically calculated the time-dependency of the diagonal element of the density matrix by nonlinear QME (NLQME) in the case where $C=1$
  up to $t_{max}=5$ ps, which was the order of the upper limit of the time region estimated as $\hbar \gamma / V_{ad}^2$ by the second-order perturbation. Fig. \ref{fig:result-fig} shows the results by NLQME, linear QME (LQME) for the case of $C=0$,
  and hierarchical equation (HEOM)\cite{Ishizaki2009}.
  Indeed there were some differences between the results by NLQME (LQME) and HEOM in shorter intervals of time ($t < 1$ ps) due to the approximation by the Markovian limit. However, the results by NLQME were similar to those by HEOM when $t > 1$ ps.
  
  To numerically analyze the energy gap $ E_d - E_a $, the coupling strength $V_{ad}$, and the reorganization energy $\lambda$ dependencies of the diagonal density matrix element $r_{dd}(t)$ by NLQME, LQME, and HEOM,
  we define the average of difference $ \Delta _ {{\rm L - HEOM}} $ ($ \Delta _ {{\rm nonL - HEOM}} $) between the results by NLQME (LQME) and those by HEOM as:
\begin{align}
\Delta_{{\rm L-HEOM}} &\equiv \sqrt{\frac{1}{t_{\rm max}}\int_0^{t_{\rm max}} ds (r^{{\rm L}}_{dd}(s)-r^{\rm HEOM}_{dd}(s))^2}
                        ,
                        \label{eq:Delta_L-H}
  \\
\Delta_{{\rm nonL-HEOM}} &\equiv \sqrt{\frac{1}{t_{\rm max}}\int_0^{t_{\rm max}} ds (r^{{\rm nonL}}_{dd}(s)-r^{\rm HEOM}_{dd}(s))^2}
,
                        \label{eq:Delta_nL-H}
\end{align}
where $r^{{\rm nonL}}_{dd} (s) $ is the diagonal density matrix element calculated by NLQME and $ r^{\rm L}_{dd} (s) $ is that by LQME, and
$ r^{\rm HEOM} _ {dd} (s) $ is the diagonal element of the reduced density matrix by HEOM.

\begin{figure}
  \centering
  \includegraphics[width=7cm]{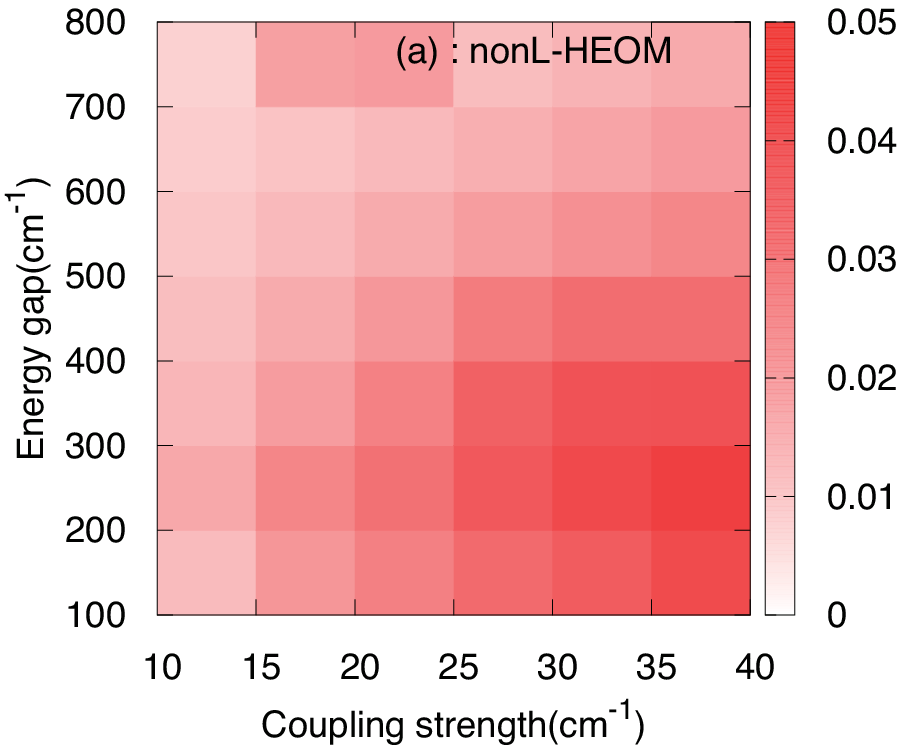}
  \includegraphics[width=7cm]{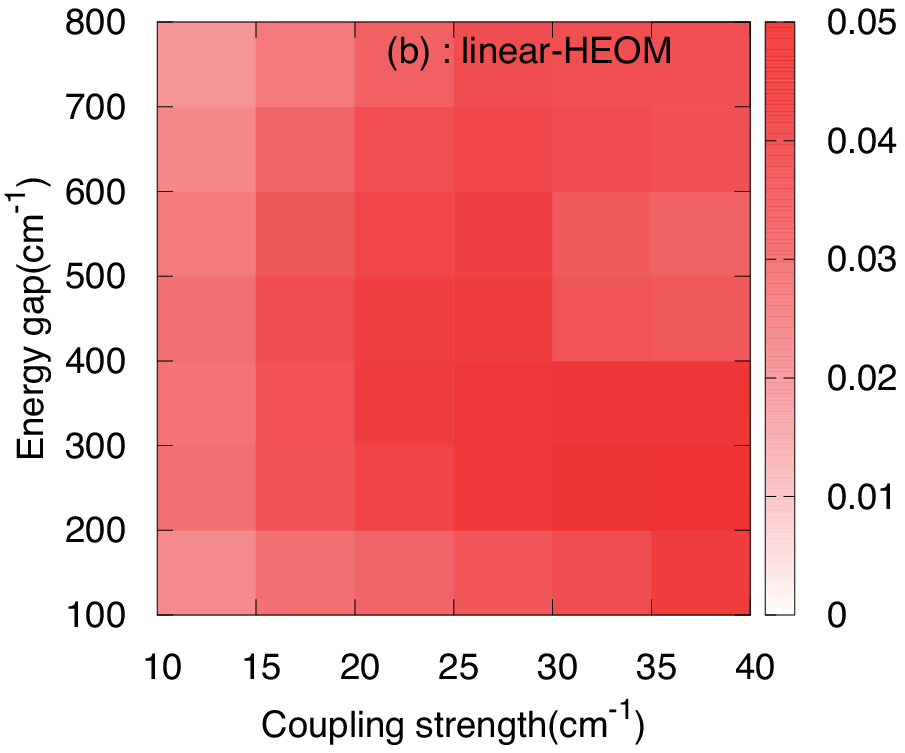}
\\
\caption{$V_{ad}$ and $E_d-E_a$ dependency of (a) $\Delta_{\rm nonL-HEOM}$ and (b) $\Delta_{\rm L-HEOM}$. The other parameters are $\lambda = 200$ cm$^{-1}$, $\gamma = 500$ cm$^{-1}$, $T = 300$ K. }
\label{fig2}
\end{figure}

The results of the coupling strength $V_{ad}$ and the energy gap $E_d - E_a$ dependencies of the average differences $\Delta_{{\rm nonL-HEOM}}$ and $\Delta_{{\rm L-HEOM}}$ are shown in Fig. \ref{fig2}. 
In the region where the energy gap is considerable, the NLQME results match the HEOM results relatively well. 
In the region of small $ V_{ad} $, which corresponds to the applicable region from F\"orster theory, the results by NLQME are more accurate than the results by LQME. Hence, the nonlinear term in NLQME plays an important role in such a case.
On the other hand, as the value of $V_{ad}$ increases, both results by LQME and NLQME fail to match the HEOM results. 
However, the value of $\Delta_{{\rm nonL-HEOM}}$ is smaller than that of $\Delta_{{\rm L-HEOM}}$. 
This implies that the applicable parameter region for NLQME is more widened than that for LQME.
In other words, since the applicability by NLQME improves in the direction of increasing $ E_d - E_a$ and $V_{ad} $, NLQME can extend the applicability of the theory from the limit of F\"orster theory to the intermediate coupling region.

\begin{figure}
\centering
\includegraphics[width=7cm]{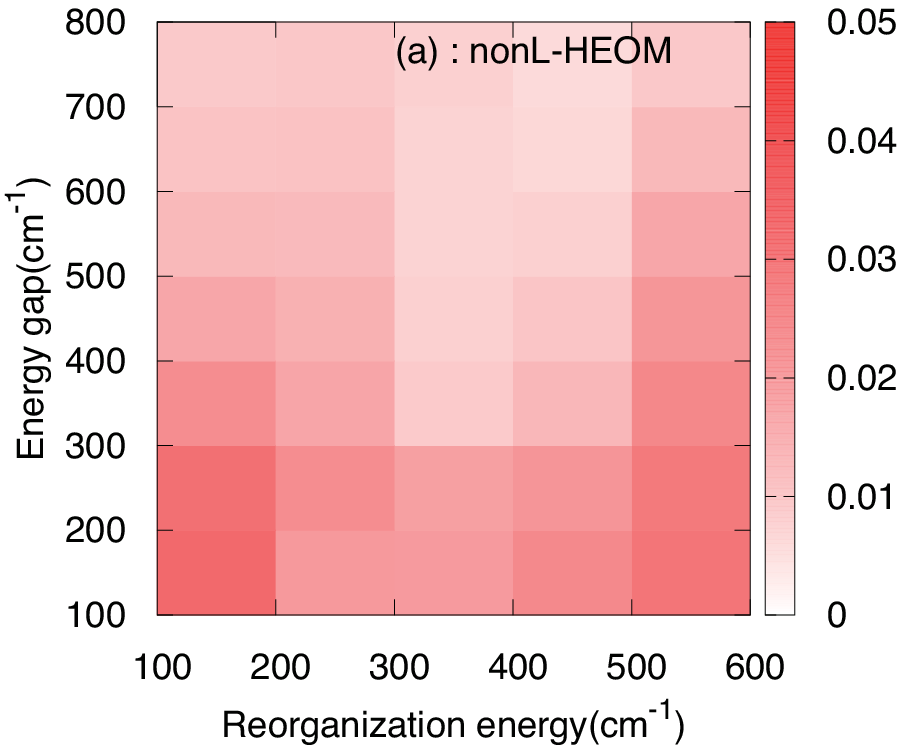}
\includegraphics[width=7cm]{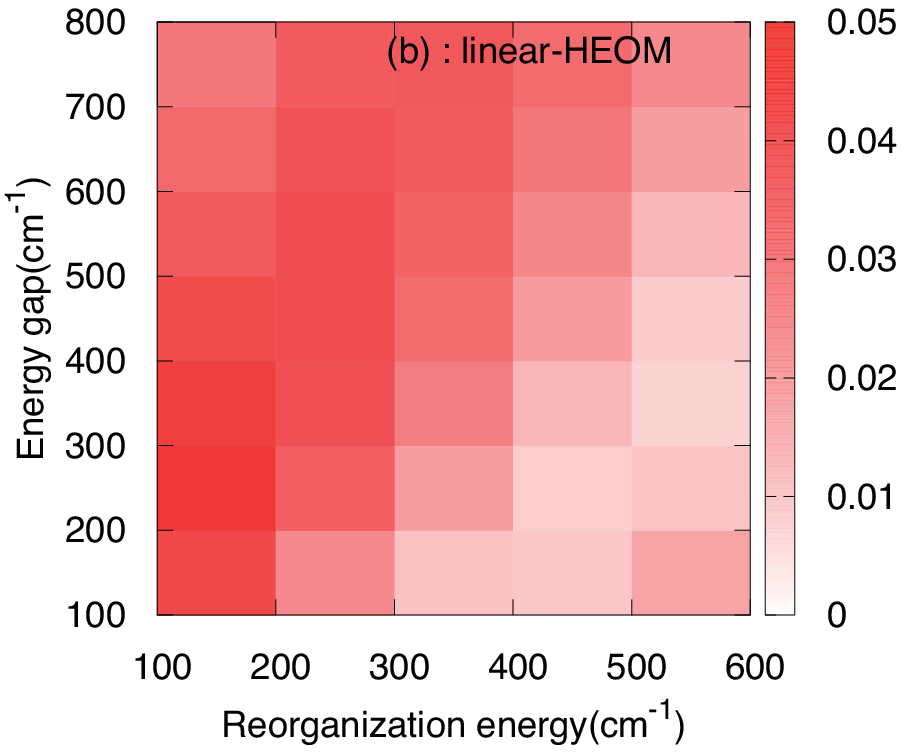}
  \caption{$\lambda$ and $E_d - E_a$ dependency of (a) $\Delta_{\rm nonL-HEOM}$ and (b) $\Delta_{\rm L-HEOM}$. The other parameters are $V_{12} = 20$ cm$^{-1}$, $\gamma = 500$ cm$^{-1}$, $T = 300$ K. }
\label{fig3}
\end{figure}

Fig. \ref{fig3} shows the reorganization energy $ \lambda $ and the energy gap $E_d - E_a$ dependencies of the average differences. 
The reproducibility of NLQME with respect to the reorganization energy is better than that of LQME when the energy gap is more than approximately 400 cm$^{-1}$. However, there is a region where the results by LQME are better than those by NLQME---namely, when the reorganization energy $\lambda$ is over 400 cm$^{-1}$ and the energy gap $E_d - E_a$ is smaller than $\lambda$.

\section{Discussion}

At present, many theoretical studies employing QME are based on conventional projection operators, for which the dynamics of the irrelevant system must be in strong thermal equilibrium regardless of the state of the relevant system. 
Here, we focused on the dynamics of the relevant system interacting with the irrelevant system regardless of the equilibrium state,
because the time-dependent projection operator techniques provide nonequilibrium dynamics in the irrelevant system.

We applied time-dependent renormalization to the conventional nonlinear QME.
Although Linden and May also derived a time convolution 
QME using the time-dependent projection operator by Willis and Picard, 
  they found that the nonlinear terms in the derived equation cancel identically
in the Markov approximation\cite{Linden1998}.
Hence, their equation coincides with the conventional LQME.
This cancellation can be attributed to simple perturbative expansion,
corresponding to the case of $C=0$ in Eqs. (\ref{eq:final-general-express}) or (\ref{eq:final-markovian-express}) in our theory.
By contrast, our proposed formula has a new nonlinear term
and the derived nonlinear terms do not cancel 
in the Markov approximation.
These differences could be due to the renormalization in the case of $C\neq 0$ in Eqs. (\ref{eq:final-general-express}) or (\ref{eq:final-markovian-express}) in our theory.
Recently, the formalism for a closed equation of the reduced density operator based on the self-consistent Mori projector technique (c-MoP) was offered by Degenfeld-Schonburg and Hartmann\cite{PhysRevB.89.245108}.
Their numerical analysis in the stationary state demonstrated highly accurate results with the c-MoP comparable with mean field theory.
  The c-MoP derives closed nonlinear quantum master equations for the N-body system using traditional projection operators.
  Derivation of the equation involves focusing on
  each element in the reduced density matrix. Specifically, the derived equations relating to the relevant system are represented exclusively by each
  element of the reduced density matrix.
  By assuming translation invariance by the symmetry of lattice,
  the derived QME was reduced to equations for cluster of subsystems.
  In the following, to compare our method by the time-dependent renormalization to that by the c-MoP or by Willis and Picard, we briefly summarize the approximations in this paper. 

The starting point is the time-convolutionless equations (\ref{eq:exact_r}) and (\ref{eq:exact_R}).
We applied time-dependent renormalization $H^{0r}(t)$ and $H'^{r}(t)$ as Eqs. (\ref{eq:H0r}) and (\ref{eq:H'r}) to these equations in section \ref{section:3.1}.
In deriving Eq. (\ref{eq:eq_of_r-29}) from the renormalized time convolutionless Eq. (\ref{eq:exact_renorm_r-2}), we performed second-order perturbative expansion with renormalized perturbative term $H'^r(t)$, which corresponds to Born approximation.
The first term on the r.h.s of Eq. (\ref{eq:Approx_RI}) which approximates the reduced density matrix of the irrelevant system $R_{\mathcal{I}}(t)$ with first order perturbation, corresponds to that of the initial condition $R_{\mathcal{I}}(0)$ by Born approximation, and the second term indicates that the dynamics of the relevant system is considered.
Since the effect of the second term on the r.h.s of Eq. (\ref{eq:Approx_RI}) is assumed to be perturbatively weak, it causes small fluctuations around the thermal equilibrium state.
One of the reduced density matrices of the relevant system $r_{\mathcal{I}}(s)$ in the second and third terms of the right-hand side of Eq. 
(\ref{eq:final-general-express}) depends on the time variable $s$ indicating non-Markovian character.
These originate because $r_{\mathcal{I}}(s)$ of the second term is due to the time-dependent projection operator $Q(s)$ in Eq. (\ref{eq:eq_of_r-29}), and
$r_{\mathcal{I}}(s)$ of the third term occurs under the influence of the dynamics in the irrelevant system of Eq. (\ref{eq:Approx_RI}).
By using Markov approximation, $r_{\mathcal{I}}(s)$ can be rewritten as $r_{\mathcal{I}}(t)$ and then the equation is approximately expressed as Eq. (\ref{eq:final-markovian-express}).
In the Markov limit, the final formula is expressed as Eq. (\ref{eq:final-Born-markovian-express}), where the effect of renormalization is included on the third term on the r.h.s of this equation.
The first and second terms of the r.h.s of Eq. (\ref{eq:final-Born-markovian-express}) are the same as the conventional linear QME.
The time correlation function in the third term of the r.h.s of Eq. (\ref{eq:final-Born-markovian-express}) is expressed similar to that in the second term (see Appendix).
Therefore, timescales $\tau_B$ and $\tau_R$ in the Born-Markov approximation ($\tau_B \ll \tau_R$) applied in Eq. (\ref{eq:final-Born-markovian-express}) are the correlation time of the bath ($\tau_B \sim \hbar / \gamma$ in this study) and the relaxation time of the relevant system ($\tau_R \sim 1/k_{ad}$ in this study) when approximately derived by the conventional linear QME.
The final form of our QME to which the secular approximation ($ \tau_S \sim \hbar/|E_d - E_a| \ll \tau_R $ in this study) is applied for numerical analysis is expressed by Eqs. (\ref{eq:nlqme-for-forster1}) and (\ref{eq:nlqme-for-forster2}).
Our QME is expressed as closed formula only by the reduced density matrix of the relevant system.
The nonlinear terms that include the coefficient $C$ in our QME are new terms yielded by the time-dependent renormalization procedure in this study,
while the new coefficient $C$ does not appear in the methods by Willis and Picard, and by the c-MoP.
In the c-MoP method, Born approximation is applied in the derivation of the QME.
In the method by Willis and Picard, the Born approximation, the Born-Markov approximation and the approximation of the Markov limit are stepwise applied in the derivation of each parallel QME
for the relevant and irrelevant systems.
In this study, at each stage of approximations summarized above, approximation is performed so that the contribution of the coefficient $C$ included in the renormalized perturbation term remains in the lowest order.
Depending on the value of the coefficient $C$, the non-linear terms in our QME contribute to correct the conventional linear QME in the relevant system through the dynamics in the irrelevant system.
This is not included in the methods by the c-MoP, and by Willis and
Picard, and is a new effect added by this study.

In this paper, through the renormalization technique in which the dynamics of the irrelevant system are renormalized into the dynamics of the relevant system, the renormalized irrelevant system appears to achieve thermal equilibrium in a mathematical sense.
In addition, the strength of the renormalized perturbative terms $H'^{r}$ should be less than that of the original perturbation terms $H'$. Hence, the applicable timescale can be extended beyond that without the renormalization technique.
The conspicuous difference between the numerical analysis of our proposed formalism and that of the conventional analysis appears when the timescale is much larger.

  The inset of Fig. \ref{fig:result-fig} shows the numerical results over a short period of time. It can be seen that the initial condition for numerical analysis with our theory is same as that by HEOM. With HEOM, however, the probability immediately following $t=0$ increases slightly, because the process is non-Markovian over a short period of time.
  In this study, we assumed that the value of $\gamma$ is large. This means that the relaxation time is very fast. When the value of $\gamma$ is small, the reaction process is non-Markovian, and the upper limit of the time integration $t_{max}$ in Eqs. (\ref{eq:Delta_L-H}) and (\ref{eq:Delta_nL-H}) is small. In such a case, since the values of Eqs. (\ref{eq:Delta_L-H}) and (\ref{eq:Delta_nL-H}) are large, quantitative discrepancies appear between the numerical results by our theory and those by HEOM, as shown in Figs. \ref{fig2} and \ref{fig3}.

The numerical results in Figs. \ref{fig2} and \ref{fig3} show that the proposed theory is strongly dependent on the size of the energy gap.
When the size of the energy gap is smaller than the thermal energy, with $|E_d - E_a| < k_B T$, the state immediately after transition at site $|d\rangle$ $\to$ site $|a\rangle$ in the relevant system is near the thermal equilibrium state for the irrelevant system without renormalization.
In such a case, the dynamics for the irrelevant system may be applicable as a thermal bath. However, in the opposite case, where $|E_d - E_a| > k_BT$, the state immediately after transition in the relevant system passes far from the thermal equilibrium state for the irrelevant system without renormalization.
In such situations, insofar as there is considerable effective relaxation time in the environment before attaining the steady state, 
the dynamics of the environment might be far from the equilibrium state immediately after the occurrence of a state transition.
Hence, the perturbative approach using the conventional projection operator is inapplicable in such a situation.
The effect of renormalization in this study should account for the feedback of such nonequilibrium states on the dynamics of the irrelevant system.

However, there were some quantitative discrepancies between the results of the proposed theory and those of the HEOM.
This may be due to the delocalization effect caused by the small energy gap, because the electronic state approaches resonance in such cases.
In this paper, our numerical analysis considered the transfer process in terms of the theory for the weak electronic coupling case exclusively. We may need to analyze the dynamics in the opposite limiting case as well, that is,
using electron-phonon coupling
as the perturbation term. The resulting QME could be applied to the relaxation processes in the delocalized exciton states.
To develop our theory more generally, the discrepancies between it and HEOM could be corrected using the perturbative approach in the modified Redfield theory \cite{Zhang1998,Yang2002}, which reproduces F\"orster theory and Redfield theory for the limiting cases.
Recently, a coherent modified Redfield theory \cite{HwangFu201546,:/content/aip/journal/jcp/142/3/10.1063/1.4905721} was constructed using a pure dephasing reference system master-equation method \cite{Golosov2001} based on the modified Redfield theory. However, to use this modified Redfield theory,
we may need to derive a more exact nonlinear QME in the intermediate coupling case.

  After the analysis shown in Fig. \ref{fig3}, it is clear that the theory by nonlinear QME, namely for the case of $C=1$ in Eq. (\ref{eq:final-markovian-express}), is difficult to apply in all parameter areas,
  where it holds that the energy gap $E_d - E_a$ is smaller than the reorganization energy $\lambda$.
  In this parameter region, it is evident that the linear master equation for the case of $C=0$ in Eq. (\ref{eq:final-markovian-express}) achieves better results than the nonlinear formalism.
  This shows that 
  there may be a relationship between the parameter region and the variable $C$.
  In other words, the degree of the feedback effect from the irrelevant system changes depending on the parameter region.
  In order to include this effect, 
  it is necessary to introduce a formalism that adjusts the variable $C$ according to the parameter region.

We previously researched the application of a variational principle to QME (which results in a so-called variational master equation) \cite{Fujihashi2014,Kimura2014}.
According to this approach,
the free energy function is used to obtain an optimized trial function for the variational method.
The variational parameters in the variational master equation should generally vary as functions of time, although the optimized variational parameter is applicable to the thermal equilibrium state.
Although some trial functions for the polaron problem have been proposed \cite{McCutcheon2011,:/content/aip/journal/jcp/140/24/10.1063/1.4884275,doi:10.1021/jp1093492,C5CP06871A}, from the point of view of the polaron transformation, we suppose that the time-dependent renormalization in this paper corresponds 
to a polaron 
transformation. 
Specifically, our numerical analysis was limited exclusively to strong constraints in the case of $C=1$ for the NLQME.
If weaker conditions could be imposed
by including the variable $C$ as a variational parameter, these problems might be partially or fully overcome.

The numerical results for the proposed theory show that this theoretical approach can be applied over large timescales. For example, in a photosynthetic antenna and its reaction center, a fast light-harvesting process occurs, followed by a slower electron transfer reaction.
In the leading studies in this field, attempts have been made to understand the mechanism that produces electron transfer reactions after excitation energy is transferred to the reaction center \cite{Raszewski2008,Shibata2013}.
Each of these elementary reactions has an intrinsic timescale, and they are hierarchically different from each other.
The proposed theory could apply to 
cases in which the thermalization process produced by the nonequilibrium dynamics of the environment has an important influence on the reactions.

Finally, the only modification to the approximated final form in the proposed theory is the inclusion of the nonlinear term in Eq. (\ref{eq:final-general-express}).
Concerning the numerical analysis, such a modification is easy to implement in the source program, and the computational cost is almost the same as that for conventional QME.
Hence, the proposed theory offers advantages for the analysis of reaction dynamics over large timescales.

\section{Conclusion}

In this paper, a new QME was constructed by applying time-dependent renormalization to the
nonlinear QME
derived by Shibata and Hashitsume\cite{Shibata1979}. 
The general expression of the proposed equation, 
which might account for the influence of dynamic environmental feedback on the irrelevant system,
has nonlinear terms in addition to the conventional linear QME.
As an example, 
the transfer dynamics in relevant two-site systems with a nuclear vibrational environment based on the conventional theory for the weak electronic coupling case were numerically analyzed in a comprehensive parameter region.
We found that, when the energy gap is large,
the proposed theory reproduces the results given by HEOM, 
  even when the differences between the conventional theory and HEOM are considerable.
  Consequently, the proposed theory offers significant advantages when studying the dynamics of a system over large timescales, 
  particularly when the thermalization caused by the non-equilibrium dynamics of the environment has an important influence on the reactions.

\appendix

\section{Kernel calculation}

We first approximate $\mbox{Tr}_f[L_I'(t) r_{\mathcal{I}}(t) \rho_B]$ as zero by random phase approximation.
The approximated nonlinear QME is expressed as follows:
\begin{align}
  &
  \frac{\partial r_{\mathcal{I}}(t)}{\partial t} 
    =
    -
    \frac{1}{\hbar^2}
    \int_0^{ \infty } ds
    \mbox{Tr}_f
    [
    L_I'(t)
    L_I'( t-s)
    r_{\mathcal{I}}(t) \rho_B
    ]
    \notag \\
  &
    +
    \frac{C}{\hbar^2}
    \int_0^{ \infty } ds
    \mbox{Tr}_f
    [
    L_I'(t)
    r_{\mathcal{I}}(t)
    \mbox{Tr}_m
    [
    L_I'( t-s )
    r_{\mathcal{I}}(t) \rho_B
    ]]
    .
    \label{eq:NLQME_for_redfield}
\end{align}
Here, we derive the formalisms of the kernel in the integrand of Eq. (\ref{eq:NLQME_for_redfield}).

First, we analyze the kernel from the first term of the r.h.s. of Eq. (\ref{eq:NLQME_for_redfield}). The kernel by the Liouville operator is re-expressed as the commutator as follows:
\begin{align}
    \mbox{Tr}_f
    [
    L_I'(t)
    L_I'( t' )
    r_{\mathcal{I}}(t) \rho_B
    ]
  =
    \mbox{Tr}_f
    [
  [
  H'_I(t)
  ,
  [
  H'_I(t')
    ,r_{\mathcal{I}}(t) \rho_B
  ]
  ]
  ]
  .
  \label{eq:A4}
\end{align}
Expanding Eq. (\ref{eq:A4}) and taking a matrix element of $\langle i|$ and $|j\rangle$, we obtain
\begin{align}
  &
    \langle i|
  \mbox{Tr}_f
    [
  [
  H'_I(t)
  ,
  [
  H'_I( t')
    ,r_{\mathcal{I}}(t) \rho_B
  ]
  ]
  ]
  |j \rangle
    \notag \\
  &
    =
    \sum_{k,l}
    [
    \langle
    H'_{ik}(t)
    H'_{kl}(t' )
    \rangle_B
    r_{lj}(t) 
    -
    \langle
    H'_{lj}(t' )
    H'_{ik}(t)
    \rangle_B
    r_{kl}(t) 
    \notag \\
  &
    -
    \langle
    H'_{lj}(t)
    H'_{ik}( t' )
    \rangle_B
    r_{kl}(t) 
    +
    \langle
    H'_{kl}( t' )
    H'_{lj}(t)
    \rangle_B
    r_{ik}(t) 
    ]
    ,
    \label{eq:A5}
\end{align}
where we define the bracket $\langle \cdots \rangle_B$ as $\mbox{Tr}_f[\cdots \rho_B]$, $\langle i|H'_I(t)|j\rangle$ as $H'_{ij}(t)$, and $\langle i| r_{\mathcal{I}}(t) |j\rangle$ as $r_{ij}(t)$.

Second, the kernel from the second term on the r.h.s of Eq. (\ref{eq:NLQME_for_redfield}) is expressed by the commutator as
\begin{align}
  &
    \mbox{Tr}_f
    [
    L_I'(t)
    r_{\mathcal{I}}(t)
    \mbox{Tr}_m
  [
    L_I'( t' )
  r_{\mathcal{I}}(t) \rho_B
  ]
    ]
    \notag \\
  &
    =
    \mbox{Tr}_f
  [
  [
    H'_I(t)
  ,
  r_{\mathcal{I}}(t)
  \mbox{Tr}_m
  [
  [
  H'_I( t' )
  ,
  r_{\mathcal{I}}(t) \rho_B
  ]
  ]
  ]
    ]
    .
  \label{eq:A6}
\end{align}
In the same way as Eq. (\ref{eq:A5}), the matrix element of $\langle i|$ and $|j\rangle$ of Eq. (\ref{eq:A6}) is expressed as
\begin{align}
&
  \langle i|
  \mbox{Tr}_f
  [
  [
    H'_I(t)
  ,
  r_{\mathcal{I}}(t)
  \mbox{Tr}_m
  [
  [
  H'_I( t' )
  ,
  r_{\mathcal{I}}(t) \rho_B
  ]
  ]
  ]
  ]
  |j \rangle
    \notag \\
  &
        =
    \sum_{k,l,m}
  [
    (
    \langle
    H'_{il}(t)
    H'_{km}( t' )
    \rangle_B
    r_{mk}(t)
    -
    \langle
    H'_{mk}( t' )
    H'_{il}(t)
    \rangle_B
    r_{km}(t)
    )
    r_{lj}(t)
    \notag \\
  &
    -
    (
    \langle
    H'_{lj}(t)
    H'_{km}( t' )
    \rangle_B
    r_{mk}(t) 
    -
    \langle
    H'_{mk}( t' )
    H'_{lj}(t)
    \rangle_B
    r_{km}(t) 
    )
    r_{il}(t)
    ]
    .
\end{align}

Let us define the following function as
\begin{align}
  \kappa_{ilkm}(t)
  =
  \frac{1}{\hbar^2}
  \int_0^{ \infty}
  ds
  \langle
  H'_{il}(t)
  H'_{km}( t-s )
  \rangle_B
  .
\end{align}
Inserting Hamiltonian $H^0$ in Eq. (\ref{eq:Hamiltonian-H0}) into the correlation function
$\langle H'_{il}(t) H'_{km}( t-s )\rangle_B$, it is expressed as
\begin{align}
  &
    \langle H'_{il}(t) H'_{km}( t-s )\rangle_B
  =
  V_{ad}^2
  e^{i(E_i - E_m) t /\hbar}
  e^{-i(E_l -E_k) t /\hbar}
  e^{i  (E_m - E_k)s /\hbar}
  \notag \\
  &
    \times
  \langle
  e^{iH_i^{e-p}t/\hbar}
  e^{-iH_l^{e-p}t/\hbar}
  e^{iH_k^{e-p}( t-s )/\hbar}
  e^{-iH_m^{e-p}( t-s )/\hbar}
  \rangle_B
  .
\end{align}
Hence, using the relation $\kappa_{ilkm}(t) \simeq \delta_{im} \delta_{kl} \kappa_{ikki}(t)$ by the secular approximation to ignore the correlation function when $E_i - E_m \neq E_l - E_k $ holds, we ultimately obtain the final form of the QME as follows:
\begin{align}
  &
  \frac{\partial r_{ij}(t)}{\partial t} 
    =
    -
    \sum_{k}
[
  \kappa_{ikki}(t)
    r_{ij}(t) 
    -
    \delta_{ij}
    \kappa_{kijk}^*(t)
    r_{kk}(t) 
    \notag \\
  &
    -
    \delta_{ij}
    \kappa_{kjik}(t)
    r_{kk}(t) 
    +
        \kappa_{jkkj}^*(t)
    r_{ij}(t) 
 ]
    \notag \\
  &    +
    C
                    \sum_{k}
 [
                \kappa_{ikki}(t)
    -
        \kappa_{kiik}^*(t)
    -
         \kappa_{kjjk}(t)
    +
        \kappa_{jkkj}^*(t)
]
    r_{ik}(t)
    r_{kj}(t)
    ,
    \label{eq:NLQME_for_summarized}
\end{align}
where $  \kappa_{ikki}(t)$ is expressed as
\begin{align}
  \kappa_{ikkk}(t)
  &
    =
    \frac{V_{ad}^2}{\hbar^2}
  \int_0^{ \infty }
  ds
  \langle
  e^{iH_it/\hbar}
  e^{-iH_ks/\hbar}
  e^{-iH_i( t-s )/\hbar}
  \rangle_B
  .
\end{align}

\bibliographystyle{plain}

\end{document}